%% file: acl_latex.tex
\documentclass[11pt]{article}
\usepackage[final]{acl}
\usepackage{times}
\usepackage{latexsym}
\usepackage[T1]{fontenc}
\usepackage[utf8]{inputenc}
\usepackage{microtype}
\usepackage{inconsolata}
\usepackage{graphicx}
\usepackage{amsmath}
\usepackage{amssymb}
\usepackage{booktabs}
\usepackage{multirow}
\usepackage{xcolor}
\usepackage{soul}
\usepackage{tikz}
\usepackage{pgfplots}
\usepackage{tabularx}
\usepackage{algorithm}
\usepackage{algpseudocode}
\usepackage{colortbl}
\usepackage{standalone}
\pgfplotsset{compat=1.18}
\usetikzlibrary{arrows.meta, positioning, shapes.geometric, fit, backgrounds, patterns, calc, shadows}

\definecolor{Navy}  {RGB}{ 30,  58, 138}
\definecolor{Blue}  {RGB}{ 59, 130, 246}
\definecolor{LBlue} {RGB}{219, 234, 254}
\definecolor{Slate} {RGB}{100, 116, 139}
\definecolor{Bg}    {RGB}{248, 250, 252}
\definecolor{Red}   {RGB}{220,  38,  38}
\definecolor{LRed}  {RGB}{254, 226, 226}
\definecolor{Green} {RGB}{ 22, 163,  74}
\definecolor{LGreen}{RGB}{220, 252, 231}

\tikzset{
  base box/.style={
    rectangle, rounded corners=6pt,
    minimum width=3.2cm, minimum height=0.85cm,
    font=\small\sffamily, align=center,
    drop shadow={opacity=0.08, shadow xshift=1.5pt, shadow yshift=-1.5pt}
  },
  box/.style={base box, draw=Blue, fill=LBlue, line width=0.8pt, text=Navy},
  hbox/.style={base box, draw=Navy, fill=Navy, line width=0.9pt, text=white, font=\small\bfseries\sffamily},
  lbox/.style={base box, draw=Red!70, fill=LRed, line width=0.8pt, text=Red!80!black},
  obox/.style={base box, draw=Green!70!black, fill=LGreen, line width=0.8pt, text=Green!50!black},
  tag/.style={
    rectangle, rounded corners=4pt, inner xsep=6pt, inner ysep=3pt,
    fill=#1, font=\scriptsize\bfseries\sffamily, text=white,
    drop shadow={opacity=0.15, shadow xshift=1pt, shadow yshift=-1pt}
  },
  arr/.style={-{Stealth[length=6pt,width=4.5pt]}, line width=0.9pt, color=Blue},
  darr/.style={-{Stealth[length=5pt,width=4pt]}, line width=0.8pt, color=Slate, dashed},
}

\definecolor{ourblue}{RGB}{219,234,254}   % light blue
\definecolor{ourborder}{RGB}{59,130,246}  % darker blue for rules

\title{Test-Time Training for Zero-Resource Dense Retrieval Reranking}

\author{
  Shiyan Liu$^{1}$ \quad Yichen Li$^{2}$ \\
  $^{1}$Huazhong University of Science and Technology \quad $^{2}$ByteDance \\
  \texttt{shyl@hust.edu.cn} \quad \texttt{liyichen.1@bytedance.com}
}

\begin{document}
\maketitle
%% ============================================================
\begin{abstract}
%% ============================================================
Dense retrievers excel at first-stage candidate generation but lack
effective reranking in zero-resource settings. Existing approaches face a fundamental dilemma: cross-encoders deliver strong reranking quality but require costly supervised training and incur high latency, while unsupervised BM25 reranking consistently degrades dense retrieval performance on most of BEIR benchmarks. We propose \textbf{DART} (\textbf{D}ense \textbf{A}daptive \textbf{R}eranking at \textbf{T}est-time), which resolves this dilemma by adapting the scoring function at inference time. For each query, the top-ranked documents serve as pseudo-positive examples and the bottom-ranked as pseudo-negative examples, providing noisy but readily available supervision to adapt a bilinear scoring matrix $W$ via a small number of gradient updates. We further introduce a confidence-weighted margin loss and a cross-query momentum buffer that warm-starts adaptation across queries. On six BEIR benchmarks, DART achieves a mean per-dataset relative NDCG@10 gain of \textbf{+2.1\%} over the dense retrieval baseline with under 10ms additional latency per query, demonstrating a powerful capability for zero-shot performance enhancement and cross-domain generalization.
\end{abstract}

\section{Introduction}
\label{sec:intro}

The modern information retrieval pipeline is typically organized as
a two-stage cascade: a fast first-stage retriever narrows the corpus
to a candidate set, which is then reranked by a more precise but
computationally expensive model \citep{lin2022pretrained,guo2020deep}.
Bi-encoder dense retrievers \citep{karpukhin2020dense,reimers2019sentence} have become the standard first stage,
offering strong recall with sub-millisecond per-document scoring.
However, reranking remains an open problem in \emph{zero-resource}
deployments where no labeled relevance judgments exist for the target
domain.

Supervised rerankers address this with extra training. Cross-encoders
\citep{nogueira2019passage,nogueira2020document} jointly attend to
query and document, achieving high accuracy at the cost of 200--500ms
latency and substantial labeled data.
Recent LLM-based rerankers \citep{sun2023chatgpt,weller2025rank1}
push accuracy further but amplify both requirements.
In the absence of training data, practitioners typically fall back to
the dense retrieval ranking itself---forgoing any reranking step
entirely---because no lightweight, reliable alternative exists.
This is especially true in deployments built entirely around vector
databases \citep{johnson2019billion}, where only dense embeddings are
indexed and lexical systems such as BM25 \citep{robertson2009probabilistic} are not available.

We observe that a useful supervision signal is already present at
inference time, without any external resource: the ranked list
produced by the dense retriever itself.
The top-ranked documents for a given query are likely relevant; the
bottom-ranked are likely not.
Although this pseudo-labeling is noisy, it captures query-specific
relevance structure that the fixed, query-agnostic cosine scoring
function cannot exploit.
This motivates a \emph{Test-Time Training} (TTT) approach
\citep{sun2020test,liu2021ttt++}: rather than modifying query or
document representations, we adapt the \emph{scoring function}
directly for each incoming query using only its own retrieved
candidates as supervision.

We propose \textbf{DART}, which frames reranking as a per-query
optimization problem. Given a query, we initialize a bilinear scoring matrix $W$ to the identity and perform a small number of gradient steps using a confidence-weighted margin loss over pseudo-labeled positives and negatives drawn from the top-$K$ retrieved documents. We additionally introduce a cross-query momentum buffer that accumulates adaptation signals across the query stream to warm-start each new query, and a dataset-adaptive optimizer selection strategy that balances convergence speed against pseudo-label noise. Evaluated on six BEIR benchmarks \citep{thakur2021beir}, DART achieves a mean per-dataset relative NDCG@10 gain of +2.1\% over the dense retrieval baseline with under 10ms latency per query.

Our contributions are summarized as follows:
\begin{itemize}
  \item We propose DART, a principled TTT framework for
        zero-resource dense retrieval reranking that adapts a bilinear
        scoring matrix at inference time using pseudo-labels derived
        directly from the dense retrieval ranking, requiring no external
        resource.
  \item We empirically demonstrate that DART improves over the dense
        retrieval baseline on five of six BEIR benchmarks with a mean
        per-dataset relative NDCG@10 gain of $+2.1\%$ and under 10ms
        additional latency per query.
  \item We provide interpretability analysis showing that $W$ updates
        concentrate in a low-dimensional subspace correlated with
        query difficulty, providing empirical evidence for the structural basis of
        cross-domain generalization.
\end{itemize}

%% ============================================================
\section{Related Work}
\label{sec:related}
%% ============================================================

\subsection{Neural Reranking}

Neural reranking has evolved through three generations.
Early cross-encoder models \citep{nogueira2019passage} apply BERT
\citep{devlin2019bert} to jointly encode query-document pairs,
achieving strong performance at the cost of high latency.
MonoT5 \citep{nogueira2020document} reformulates reranking as a
sequence-to-sequence generation task.
ColBERTv2 \citep{santhanam2022colbertv2} introduces late
interaction to balance effectiveness and efficiency.
More recently, listwise Large Language Model (LLM) rerankers \citep{sun2023chatgpt,
pradeep2023rankvicuna} leverage the in-context learning
capabilities of LLMs.
\citet{weller2025rank1} train rerankers on reasoning traces from
DeepSeek-R1 \citep{DeepSeekAI2025DeepSeekR1IR}, achieving state-of-the-art performance by exploiting
test-time compute in the form of chain-of-thought reasoning---a
complementary direction to ours, which targets lightweight parameter
adaptation rather than extended generation.
All supervised rerankers require labeled training data, limiting
applicability in zero-resource domains.

\subsection{Unsupervised Domain Adaptation}
\label{sec:unsup}

GPL \citep{wang2022gpl} generates pseudo training pairs using a
cross-encoder teacher for unsupervised domain adaptation, but still
requires offline training.
AugTriever \citep{zhuang2023augmenting} constructs pseudo query-document
pairs via query extraction and generation for unsupervised retrieval
pretraining.
\citet{meng2022unsupervised} propose relevance-aware contrastive
pretraining that weights pseudo-positive pairs by estimated relevance,
improving Contriever \citep{Izacard2021UnsupervisedDI} on BEIR \citep{thakur2021beir} without labeled data.
UDAPDR \citep{saad2023udapdr} uses LLMs to generate
domain-specific queries for zero-shot dense retrieval adaptation.
These methods improve the retrieval model itself through data
augmentation and pretraining; DART instead adapts the
\emph{scoring function} at inference time with no offline training.

\begin{figure*}[!t]
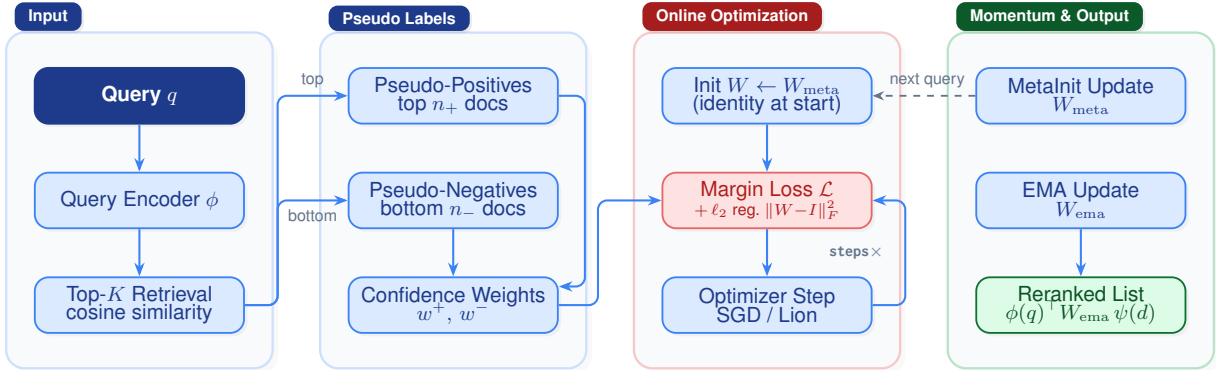

\centering
% \textwidth 指双栏页面的总宽度
\resizebox{1\textwidth}{!}{%
    \includestandalone{figure1}%
}
\caption{Overall DART algorithm flowchart}
\label{fig:flowchart}
\end{figure*}

\subsection{Pseudo Relevance Feedback}

Pseudo Relevance Feedback (PRF) \citep{lavrenko2017relevance} assumes
the top-$k$ retrieved documents are relevant and uses them to expand
queries.
Dense PRF methods \citep{li2023pseudo} encode feedback documents and
aggregate their embeddings with the query embedding.
ColBERT-PRF \citep{wang2023colbert} applies late interaction with
pseudo-relevant embeddings.
PromptPRF \citep{li2025pseudo} uses LLMs to extract structured
features from top-ranked documents offline, enabling small retrievers
to match larger ones.
\citet{wang2022text} show that dense retrievers benefit from
interpolation with BM25, motivating our hybrid pseudo-label strategy.
Unlike PRF methods that modify query representations,
DART modifies the \emph{scoring matrix}---a distinct and complementary
approach that preserves the original query and document embeddings.

\subsection{Test-Time Training}

TTT \citep{sun2020test} adapts model parameters at
inference time using self-supervised signals from the test input.
TTT++ \citep{liu2021ttt++} improves stability through feature alignment.
LoRA-based TTT \citep{yu2023metamath} and related work on TTT for
abstract reasoning \citep{akyurek2024surprising} demonstrate
that even a handful of gradient steps on a test instance can
substantially improve performance.
In information retrieval, TTT has not been studied as a reranking
mechanism.
The closest work is \citet{weller2025rank1}, which applies test-time
\emph{compute scaling} rather than test-time \emph{parameter
adaptation}.
DART is among the first methods to explore parameter-level TTT specifically for retrieval reranking.

\section{Method}
\label{sec:method}

\subsection{Setup}
\label{setup}

Let $\phi: \mathcal{Q} \to \mathbb{R}^d$ and $\psi: \mathcal{D} \to \mathbb{R}^d$ be fixed, pretrained query and document encoders (e.g., a sentence transformer). Standard dense retrieval scores a query-document pair by cosine similarity after $\ell_2$-normalization:
\begin{equation}
s(q,d) = \phi(q)^\top \psi(d).
\end{equation}
This scoring function implicitly treats all embedding dimensions as equally important and independent. However, for a specific query, certain semantic dimensions are more discriminative than others. For example, a query about ``cardiovascular disease prevention'' should upweight dimensions encoding health interventions and downweight dimensions encoding economic or political concepts.

To enable query-specific reweighting of embedding dimensions, we generalize the scoring function to a bilinear form:
\begin{equation}
s_W(q,d) = \phi(q)^\top W \psi(d),
\end{equation}
where $W \in \mathbb{R}^{d \times d}$ is a transformation matrix. Decomposing $W = I + \Delta W$ reveals that the adjustment $\Delta W$ serves as a query-specific correction to the identity mapping. Initializing $W = I$ recovers the standard cosine score exactly, providing a natural starting point and a reliable baseline.

The core challenge is to estimate $\Delta W$ for each query at inference time, using only the retrieved documents as a source of noisy supervision, without any labeled data.

\subsection{Overall Framework}

We treat the reranking task as an online optimization problem. For an incoming query $q$, we first retrieve its top-$K$ documents using the initial scoring function $s(q,d)$. These top documents, albeit noisy, provide a set of pseudo-positive and pseudo-negative examples for adaptation \citep{lavrenko2017relevance}. We then update $W$ by minimizing a loss function designed to pull relevant documents closer and push irrelevant ones away, while regularizing $W$ towards the identity to avoid overfitting.

To improve the robustness and convergence of the online update, we introduce three components that mirror a standard optimization pipeline:
\begin{enumerate}
    \item \textbf{Learning objective} (Section~\ref{leob}): a loss with confidence-weighted pseudo labels and an adaptive margin.
    \item \textbf{Cross-query momentum} (Section~\ref{cqm}): temporal regularizers (MetaInit and EMA) that transfer knowledge across queries.
    \item \textbf{Optimizer selection} (Section~\ref{sec:optimizer}): a dataset-driven
    selection between a conservative optimizer (Stochastic Gradient Descent, SGD) and a more
    aggressive one (Lion \citep{chen2023symbolic}).
\end{enumerate}
The overall algorithm is illustrated in Figure~\ref{fig:flowchart}.

\subsection{Learning Objective}
\label{leob}

For a given query $q$, let the initial retrieval scores of the top-$K$ documents be $s_1 \ge s_2 \ge \dots \ge s_K$.  
We treat the top $n_{\text{pos}}$ documents as pseudo-positive and the bottom $n_{\text{neg}}$ as pseudo-negative.

\paragraph{Confidence-weighted pseudo labels.}
To reduce the impact of label noise, we assign soft weights to the pseudo-positive and pseudo-negative examples.  
Define the normalization constant for the positive set:
\begin{equation}
Z_{\text{pos}} = \sum_{j=1}^{n_{\text{pos}}} \exp(s_j / T),
\end{equation}
where $T>0$ is a temperature hyperparameter that controls the concentration of weights.  
The weight of the $i$-th pseudo-positive document is then:
\begin{equation}
w_i^{+} = \frac{\exp(s_i / T)}{Z_{\text{pos}}}, \qquad i = 1,\dots,n_{\text{pos}}.
\end{equation}

Similarly, for the pseudo-negative documents, we define:
\begin{equation}
Z_{\text{neg}} = \sum_{k=K-n_{\text{neg}}+1}^{K} \exp(-s_k / T),
\end{equation}
and the weight of the $j$-th pseudo-negative document is:
\begin{equation}
w_j^{-} = \frac{\exp(-s_j / T)}{Z_{\text{neg}}}, \qquad j = K-n_{\text{neg}}+1,\dots,K.
\end{equation}

These weights assign higher importance to examples with larger initial scores (for positives) or more negative scores (for negatives), effectively focusing the learning on high‑confidence pseudo‑labels.

\paragraph{Adaptive margin (AdaMargin).}
The loss function encourages a margin between the aggregated scores of pseudo‑positive and pseudo‑negative documents.  
Because queries vary in difficulty, we make the margin adaptive to the highest initial similarity $s_{\text{top1}} = s_1$:
\begin{equation}
\text{margin}(q) = \alpha_{\text{mar}} + \beta_{\text{mar}}\,(1 - s_{\text{top1}}),
\end{equation}
where $\alpha_{\text{mar}}$ and $\beta_{\text{mar}}$ are hyperparameters that determine the base margin and the strength of the difficulty‑based adaptation.

\paragraph{Loss.}
Recall the bilinear scoring function $s_W(q,d) = \phi(q)^\top W \psi(d)$ (Section~\ref{setup}).  
We first compute the total weighted score for the pseudo-positive documents:
\begin{equation}
P = \sum_{i=1}^{n_{\text{pos}}} w_i^{+} \, s_W(q,d_i).
\end{equation}
Similarly, the total weighted score for the pseudo-negative documents is:
\begin{equation}
N = \sum_{j=K-n_{\text{neg}}+1}^{K} w_j^{-} \, s_W(q,d_j).
\end{equation}
The ranking loss is then defined as:
\begin{equation}
\mathcal{L}_{\text{rank}} = \max\!\Bigl(0,\; \text{margin}(q) - P + N\Bigr).
\end{equation}

We add an $\ell_2$ regularization term to keep $W$ close to the identity:
\begin{equation}
\mathcal{L}_{\text{reg}} = \lambda \|W - I\|_F^2,
\end{equation}
where $\lambda>0$ is a hyperparameter.  

The total loss for the current query is:
\begin{equation}
\mathcal{L}(W) = \mathcal{L}_{\text{rank}} + \mathcal{L}_{\text{reg}}.
\end{equation}

\subsection{Cross-Query Momentum}
\label{cqm}

To transfer knowledge across queries and smooth the parameter evolution, we maintain two complementary momentum‑like states.  
Let \(t\) denote the index of the current query.  The transformation matrix after updating query \(t\) is denoted \(W_{\text{star}}^{(t)}\).

\paragraph{Meta Initialization (MetaInit).}
We learn a global initial matrix \(W_{\text{meta}}\) that is passed from one query to the next.  
Before updating query \(t\), the initial matrix is set to the meta parameter obtained after processing the previous query:
\[
W_{\text{init}}^{(t)} = W_{\text{meta}}^{(t-1)}.
\]
After obtaining \(W_{\text{star}}^{(t)}\), we update the meta parameter using the Reptile rule:
\begin{equation}\fontsize{10}{9}
W_{\text{meta}}^{(t)} = W_{\text{meta}}^{(t-1)} + \beta_{\text{meta}} \left( W_{\text{star}}^{(t)} - W_{\text{meta}}^{(t-1)} \right),
\end{equation}
where $\beta_{\text{meta}}>0$ is a meta learning rate.  
This provides an increasingly better starting point for each new query, accelerating adaptation over time.

\paragraph{Exponential Moving Average (EMA).}
We maintain an exponentially decaying average of the transformation matrices for stability:
\begin{equation}
W_{\text{ema}}^{(t)} = \alpha_{\text{ema}} \, W_{\text{ema}}^{(t-1)} + (1-\alpha_{\text{ema}}) \, W_{\text{star}}^{(t)},
\end{equation}
with $\alpha_{\text{ema}} \in (0,1)$ a decay hyperparameter.  
The smoothed matrix \(W_{\text{ema}}^{(t)}\) is used for re‑ranking the current query, which reduces the variance of the updates.

Both states are carried over across the query stream.  
MetaInit affects the initial value of the next query, while EMA smooths the output of the current query.

\subsection{Optimizer Selection}
\label{sec:optimizer}

The choice of optimizer directly affects how each query's loss is minimized and interacts with the cross-query states.  
Based on empirical observations across diverse datasets, we provide guidelines for selecting between two optimizers.

\paragraph{SGD with momentum.}
SGD with momentum ($\mu=0.9$) performs conservative updates:
\begin{equation}\fontsize{10}{9}
v_{t+1} = \mu v_t - \eta \nabla \mathcal{L}(W_t), \quad W_{t+1} = W_t + v_{t+1},
\end{equation}
where $\eta$ is the learning rate.  
This optimizer is preferable when the initial dense retrieval is noisy or the dataset suffers from high pseudo-label uncertainty (e.g., TREC-COVID, SciFact), as it avoids overfitting.

\paragraph{Lion optimizer.}
The Lion optimizer updates parameters using only the sign of the gradient:
\begin{equation}\fontsize{10}{9}
W_{t+1} = W_t - \eta \cdot \text{sign}\bigl( \beta_1 m_t + (1-\beta_1) \nabla \mathcal{L}(W_t) \bigr),
\end{equation}
with $m_t$ an exponential moving average of past gradients.  
Lion discards gradient magnitude, making it robust to scale variations and often faster to converge. It is more suitable for datasets where dense retrieval already provides clean pseudo-labels (e.g., NFCorpus, FiQA, SCIDOCS, ArguAna \citep{thakur2021beir}).

\paragraph{Practice.}
When no prior knowledge about the dataset is available, we recommend a simple warm-up adaptive strategy: process the first 50--100 queries with both optimizers, compare their average pseudo-label loss (Section~\ref{leob}), and select the optimizer with the lower loss for the remaining queries. This adds negligible overhead and eliminates manual tuning. In our experiments, we report the better result for each dataset following this rule or the empirical guidelines above.

\vspace{0.5\baselineskip}
\noindent
\textbf{Pseudo-code.} 
Algorithm~\ref{alg:dart} summarizes the complete test-time adaptation for a single query.

\begin{algorithm}[!t]
\small
\caption{DART for One Query}
\label{alg:dart}
\begin{algorithmic}[1]
\Require Query $q$; encoders $\phi, \psi$; retrieval depth $K$;
         hyperparameters $n_{\text{pos}}, n_{\text{neg}}, T, 
         \alpha_{\text{mar}}, \beta_{\text{mar}}, \lambda, 
         \alpha_{\text{ema}}, \beta_{\text{meta}}, \text{steps}, \eta$;
         optimizer (SGD / Lion)
\Ensure  Reranked list of $K$ documents
\Statex  \textit{Global state:} $W_{\!\text{meta}},\, W_{\!\text{ema}}$ 
         (initialized to $I$)

\vspace{2pt}
\State $\{(d_k, s_k)\}_{k=1}^K \gets \textsc{RetrieveTopK}(q,K)$,\quad
       $s_1 \ge s_2 \ge \cdots \ge s_K$,\quad
       $s_k = \phi(q)^\top\psi(d_k)$

\State $\mathcal{P} \gets \{d_1,\dots,d_{n_{\text{pos}}}\}$,\quad
       $\mathcal{N} \gets \{d_{K-n_{\text{neg}}+1},\dots,d_K\}$

\vspace{4pt}
\State \textbf{// Confidence weights}
\State $w_i^{+} \gets \dfrac{\exp(s_i/T)}{\sum_{i'=1}^{n_{\text{pos}}}\exp(s_{i'}/T)}$
       \quad for each $d_i \in \mathcal{P}$

\State $w_j^{-} \gets \dfrac{\exp(-s_j/T)}{\sum_{j'}\exp(-s_{j'}/T)}$
       \quad for each $d_j \in \mathcal{N}$

\vspace{4pt}
\State \textbf{// initialize transformation matrix}
\State $W \gets W_{\!\text{meta}}$

\vspace{4pt}
\State \textbf{// Online gradient updates}
\For{$t = 1$ \textbf{to} $\text{steps}$}
    \State $P \gets \displaystyle\sum_{d_i \in \mathcal{P}} w_i^{+}\,\phi(q)^\top W\,\psi(d_i)$
    \State $N \gets \displaystyle\sum_{d_j \in \mathcal{N}} w_j^{-}\,\phi(q)^\top W\,\psi(d_j)$
    \State $m \gets \alpha_{\text{mar}} + \beta_{\text{mar}}\,(1 - s_1)$
    \State $\mathcal{L} \gets \max(0,\; m - P + N) + \lambda\,\|W - I\|_F^2$
    \State $W \gets \textsc{OptimizerStep}(W,\, \nabla_W \mathcal{L},\, \eta)$
\EndFor
\State $W^{\star} \gets W$

\vspace{4pt}
\State \textbf{// Update cross-query momentum}
\State $W_{\!\text{ema}}  \gets \alpha_{\text{ema}}\, W_{\!\text{ema}}  + (1-\alpha_{\text{ema}})\, W^{\star}$
\State $W_{\!\text{meta}} \gets W_{\!\text{meta}} + \beta_{\text{meta}}\,(W^{\star} - W_{\!\text{meta}})$

\vspace{4pt}
\State \textbf{// Rerank}
\State \Return $\{d_k\}$ sorted descending by
       $s_{W_{\text{ema}}}(q,d_k) = \phi(q)^\top W_{\!\text{ema}}\,\psi(d_k)$
\end{algorithmic}
\end{algorithm}

%% ============================================================
\section{Experiments}
\label{sec:experiments}
%% ============================================================

% --------------- colour helpers ---------------
% Light blue shading for our method row; bold border rules above/below it.
% Requires \usepackage{colortbl} in preamble.

\begin{table*}[t]
\centering
\small
\setlength{\tabcolsep}{4.2pt}
\caption{%
  Results on six BEIR datasets.
  \textbf{Abbreviations:}
  NFC\,=\,NFCorpus; SCI\,=\,SCIDOCS; Argu\,=\,ArguAna; COVID\,=\,TREC-COVID.
  $^{*}$~Mean per-dataset relative change versus Dense Retrieval (BGE-small);
  negative values for supervised methods reflect training on out-of-domain data.
  $^{\dagger}$~Latency per query on NVIDIA RTX5090.
  $^{\ddagger}$~Latency estimated from the respective papers.
  $^{\S}$~ICR Avg.\ and Avg. Gain computed over 5 datasets (ArguAna excluded).
  $^{\P}$~InstUPR Avg.\ and Avg. Gain computed over 5 datasets (ArguAna excluded).
    $^{\|}$~Results for methods \emph{not} marked with this symbol are from their original papers.
  \textbf{Bold} = column-wise best among all methods.
  ``---''~=~not reported.
}
\begin{tabularx}{\textwidth}{l *{6}{>{\centering\arraybackslash}X} ccc}
\toprule
\multirow{2}{*}{\textbf{Method}}
  & \multicolumn{6}{c}{\textbf{NDCG@10 per Dataset}}
  & \multicolumn{3}{c}{\textbf{Overall}} \\
\cmidrule(lr){2-7}\cmidrule(lr){8-10}
  & \textbf{NFC} & \textbf{SCI} & \textbf{FiQA} & \textbf{Argu} & \textbf{COVID} & \textbf{SciFact}
  & \textbf{Avg.} & \textbf{Avg. Gain}$^{\,*}$ & \textbf{Latency}$^{\,\dagger}$ \\
\midrule
\multicolumn{10}{l}{\textit{Supervised Dense Retrieval}} \\
\quad ColBERT         & 0.305 & 0.145 & 0.317 & 0.233 & 0.677 & 0.671 & 0.391 & $-19.9\%$ & ---          \\
\quad DPR-MSMARCO            & 0.208 & 0.108 & 0.275 & 0.414 & 0.561 & 0.478 & 0.341 & $-31.9\%$ & ---          \\
\quad ANCE          & 0.237 & 0.122 & 0.295 & 0.415 & 0.654 & 0.507 & 0.372 & $-25.4\%$ & ---          \\
\quad MoDIR           & 0.244 & 0.124 & 0.296 & 0.418 & 0.676 & 0.502 & 0.377 & $-24.4\%$ & ---          \\
\quad TAS-B           & 0.319 & 0.149 & 0.300 & 0.427 & 0.481 & 0.643 & 0.387 & $-19.7\%$ & ---          \\
\quad RocketQAv2      & 0.293 & 0.131 & 0.302 & 0.451 & 0.675 & 0.568 & 0.403 & $-18.7\%$ & ---          \\
\quad SPLADEv2        & 0.334 & 0.158 & 0.336 & 0.479 & 0.710 & 0.693 & 0.452 & $-8.3\%$  & ---          \\
\quad ColBERTv2       & 0.338 & 0.154 & 0.356 & 0.463 & 0.738 & 0.693 & 0.457 & $-7.3\%$  & ${\sim}80$ms \\
\midrule
\multicolumn{10}{l}{\textit{Supervised Reranking}} \\
\quad MonoT5-base     & \textbf{0.378} & 0.154 & 0.376 & 0.476 & \textbf{0.796} & 0.675 & 0.476 & $-3.1\%$ & ${\sim}600$ms \\
\midrule
\multicolumn{10}{l}{\textit{Training-free Reranking}} \\
\quad Dense Retrieval (BGE-small)$^{\,\|}$ & 0.337 & 0.197 & 0.385 & 0.595 & 0.665 & 0.720 & 0.483 & ${=}0.0\%$   & ${<}1$ms     \\
\quad BM25 Rerank $^{\,\|}$    & 0.302 & 0.156 & 0.220 & 0.371 & 0.685 & 0.588 & 0.387 & $-21.2\%$ & ${<}2$ms \\
\quad ASRank          & 0.346 & 0.184 & 0.352 & 0.478 & 0.737 & 0.710 & 0.468 & $-3.8\%$  & ${\sim}200$ms$^{\,\ddagger}$ \\
\quad ICR & 0.347 & 0.171 & 0.381 & --- & 0.728 & \textbf{0.761} & $0.478^{\,\S}$ & $+0.8\%^{\,\S}$ & ${\sim}200$ms$^{\,\ddagger}$ \\
\quad InstUPR         & 0.352 & 0.190 & \textbf{0.398} & ---  & 0.730 & 0.713 & $0.477^{\,\P}$ & $+2.6\%^{\,\P}$ & ${\sim}200$ms$^{\,\ddagger}$ \\
\midrule
\multicolumn{10}{l}{\textit{Test-time Adaptation}} \\
\quad PRF-Vec ($n{=}3$)$^{\,\|}$ & 0.347 & 0.203 & 0.371 & 0.602 & 0.663 & 0.710 & 0.483 & $+0.3\%$ & ${<}2$ms \\
\quad PRF-Vec ($n{=}5$)$^{\,\|}$ & 0.341 & 0.201 & 0.362 & 0.585 & 0.671 & 0.704 & 0.477 & $-1.0\%$ & ${<}1$ms \\
% ---- our method: highlighted row ----
\rowcolor{ourblue}
\quad \textbf{DART (Ours)$^{\,\|}$}
  & \textbf{0.354} & \textbf{0.205} & 0.389 & \textbf{0.605} & 0.670 & 0.719
  & \textbf{0.490} & $+2.1\%$ & ${<}10$ms \\
\bottomrule
\end{tabularx}
\label{tab:main}
\end{table*}

\subsection{Experimental Setup}
\label{ssec:setup}

\paragraph{Datasets.}
We evaluate on six BEIR benchmark datasets \citep{thakur2021beir} spanning diverse domains:
biomedical literature (NFCorpus, SCIDOCS, SciFact), financial QA (FiQA),
argument retrieval (ArguAna), and biomedical COVID-19 retrieval (TREC-COVID).
Corpus sizes range from 3.6K to 171K documents, and dense retrieval baselines
(NDCG@10) vary from 0.197 (SCIDOCS) to 0.720 (SciFact), providing a challenging
testbed for zero-resource generalization.

\paragraph{Base retriever.}
We use BGE-small-en-v1.5 \citep{xiao2024c} (dimension $d=384$, 33M parameters)
as the fixed dense retriever. This small model ensures that improvements from
DART are not confounded by a strong base model and matches realistic
deployment constraints.

\paragraph{Baselines.}
We compare against several unsupervised or training-free methods. Dense Retrieval uses the same BGE-small encoder with cosine similarity, serving as the lower bound. PRF-Vec \citep{li2023pseudo} is a standard pseudo-relevance feedback method that averages top retrieved document embeddings. BM25 Rerank \citep{robertson2009probabilistic} reorders the dense top-100 using lexical BM25 scores, providing a purely sparse baseline. Recent training-free approaches include ASRank \citep{abdallah2025asrank}, ICR \citep{chen2024attention} (based on Llama-3.1-8B), and InstUPR \citep{huang2024instupr}. For reference, we also report numbers from supervised dense retrievers (e.g., ColBERT \citep{khattab2020colbert}, DPR-MSMARCO \citep{xin2022zero}, ANCE \citep{xiong2020approximate}, MoDIR \citep{xin2022zero}, TAS-B \citep{hofst2021efficiently}, ColBERTv2 \citep{santhanam2022colbertv2}) and the cross-encoder reranker MonoT5-base \citep{nogueira2020document}; these are not applicable in our zero-resource setting but illustrate the potential of supervised training.

\paragraph{Hyperparameters.}
All hyperparameters are fixed based on NFCorpus: $n_{\text{pos}}=5$ is the number of pseudo-positive documents, $n_{\text{neg}}=20$ is the number of pseudo-negative documents, $K=100$ is the initial retrieval depth, $T=0.1$ is the temperature for confidence weighting, $\alpha_{\text{mar}}=0.1$ and $\beta_{\text{mar}}=0.2$ are the base margin and adaptation strength, $\alpha_{\text{ema}}=0.9$ and $\beta_{\text{meta}}=0.1$ are the EMA decay rate and meta learning rate for cross-query momentum, $\lambda=10^{-3}$ is the regularization coefficient, $\text{steps}=5$ is the number of gradient updates per query, and the learning rate $\eta=10^{-2}$. The optimizer is selected per dataset following the guidelines in Section~\ref{sec:optimizer}. No dataset-specific tuning is performed.

\begin{table*}[t]
\centering
\small
\caption{Ablation study on four BEIR datasets. Gains are $\Delta$NDCG@10 relative to Dense Retrieval.}
\label{tab:ablation}
\setlength{\tabcolsep}{1.5pt}
\begin{tabular}{lcccccccc}
\toprule
\multirow{2}{*}{\textbf{Variant}} & \multicolumn{2}{c}{\textbf{NFCorpus}} & \multicolumn{2}{c}{\textbf{SCIDOCS}} & \multicolumn{2}{c}{\textbf{FiQA}} & \multicolumn{2}{c}{\textbf{ArguAna}} \\
\cmidrule(lr){2-3}\cmidrule(lr){4-5}\cmidrule(lr){6-7}\cmidrule(lr){8-9}
& \textbf{NDCG@10} & \textbf{Gain} & \textbf{NDCG@10} & \textbf{Gain} & \textbf{NDCG@10} & \textbf{Gain} & \textbf{NDCG@10} & \textbf{Gain} \\
\midrule
Dense Retrieval & 0.337 & ${=}0.0\%$ & 0.197 & ${=}0.0\%$ & 0.385 & ${=}0.0\%$ & 0.595 & ${=}0.0\%$ \\
Base online update (conf. weighting) & 0.346 & +2.7\% & 0.199 & +1.0\% & 0.363 & --5.7\% & 0.595 & 0.0\% \\
+ AdaMargin & 0.350 & +3.9\% & 0.201 & +2.0\% & 0.362 & --6.0\% & 0.595 & 0.0\% \\
+ EMA & 0.351 & +4.0\% & 0.199 & +1.0\% & 0.378 & --1.8\% & 0.596 & +0.2\% \\
+ MetaInit & 0.348 & +3.3\% & 0.197 & 0.0\% & 0.362 & --6.0\% & 0.599 & +0.7\% \\
+ EMA + AdaMargin & 0.355 & +5.3\% & 0.203 & +3.0\% & 0.378 & --1.8\% & 0.597 & +0.3\% \\
+ EMA + MetaInit & 0.349 & +3.6\% & 0.197 & 0.0\% & 0.377 & --2.1\% & 0.599 & +0.7\% \\
+ EMA + AdaMargin + MetaInit & 0.353 & +4.7\% & 0.202 & +2.5\% & 0.377 & --2.1\% & 0.605 & +1.7\% \\
% \arrayrulecolor{ourborder}
% \midrule[0.5pt]
\rowcolor{ourblue}
+ EMA + AdaMargin + MetaInit + Lion & 0.354 & +5.0\% & 0.205 & +4.1\% & 0.389 & +1.0\% & 0.605 & +1.7\% \\
% \midrule[0.5pt]
% \arrayrulecolor{black}
\bottomrule
\end{tabular}
\end{table*}

\subsection{Main Results}

DART improves over the dense retrieval baseline on five of six datasets,
achieving a mean per-dataset relative gain of $+2.1\%$ NDCG@10
(Table~\ref{tab:main}).
The largest improvement is on NFCorpus ($+5.0\%$), where the baseline is
weakest, with further notable gains on SCIDOCS ($+4.1\%$) and ArguAna
($+1.7\%$).
Modest gains are observed on FiQA ($+1.0\%$) and TREC-COVID ($+0.8\%$).
SciFact is the only dataset where DART ties the baseline ($-0.1\%$,
effectively no change), likely because the high baseline score
(0.720) leaves little headroom for unsupervised adaptation.

Compared to PRF-Vec, which degrades on FiQA and TREC-COVID and provides
near-zero average gain ($+0.3\%$ for $n{=}3$), DART delivers
consistent improvements.
BM25 Rerank is unreliable, helping only on TREC-COVID while degrading by
$-26\%$ on average across the remaining five datasets; DART
outperforms it by $+42\%$ on those five datasets.

Recent training-free LLM-based approaches (ASRank, ICR, InstUPR) show
average gains of $-3.8\%$, $+0.8\%$, and $+2.6\%$ respectively, but
require approximately 200ms per query ($\sim$20$\times$ the latency of DART). In contrast, DART runs in under 10ms per query on an NVIDIA RTX5090 GPU, making it far better suited for real-time, latency-sensitive deployment.

\begin{figure*}[t]
  \centering
  \includegraphics[width=\textwidth]{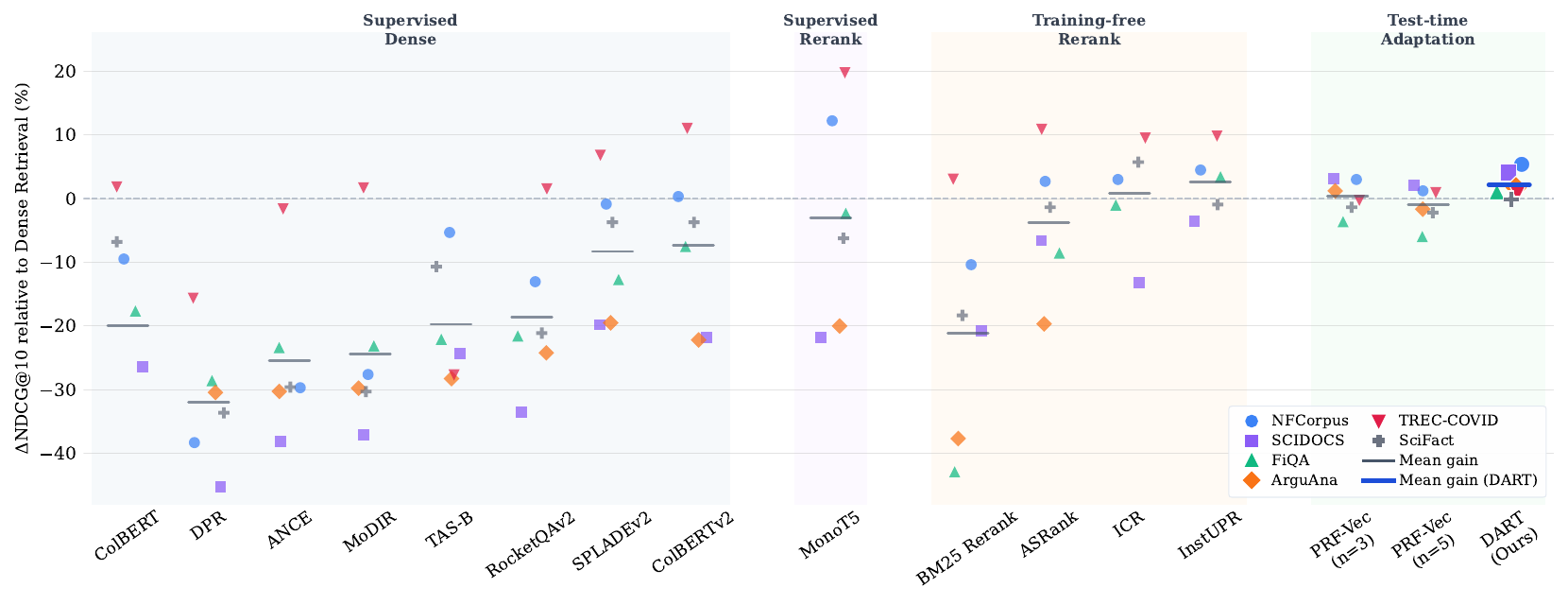}
  \caption{Per-dataset $\Delta$NDCG@10 relative to Dense Retrieval.
    Each point is one dataset; bars show the mean gain.}
  \label{fig:strip}
\end{figure*}

Notably, DART surpasses all supervised dense retrievers except
ColBERTv2 and SPLADEv2 despite using no training data. Figure~\ref{fig:strip} visualises the full gain distribution across all
methods: DART is the only training-free method with no negative outlier
on any dataset.

\begin{table}[!t]
\centering
\small
\caption{Statistics of $\|\Delta W\|_F$ on NFCorpus.}
\label{tab:update_stats}
\begin{tabular}{lccccc}
\toprule
\textbf{Statistic} & Min & 25\% & Median & 75\% & Max \\
\midrule
Value & 0.000 & 0.048 & 0.095 & 0.111 & 0.125 \\
\bottomrule
\end{tabular}
\end{table}

\subsection{Ablation Study}

We evaluate the contribution of each component on a representative subset of BEIR: NFCorpus, SCIDOCS, FiQA, and ArguAna. Table~\ref{tab:ablation} reports NDCG@10 for variants that incrementally add the modules described in Section~\ref{sec:method}, using steps=5. The base online update refers to confidence-weighted pseudo labels with a fixed margin. Gains are relative percentages over the dense retrieval baseline of each dataset.

EMA is the most universally beneficial component, yielding positive
gains on all four datasets and single-handedly recovering FiQA from
$-5.7\%$ (base) to $-1.8\%$; this aligns with the query difficulty
analysis (Section~\ref{analysis-query}), which shows that easy queries
(high $s_{\mathrm{top1}}$) benefit most from smoothing rather than
aggressive per-query adaptation.
AdaMargin contributes most on NFCorpus, where the wide spread of
query difficulty ($s_{\mathrm{top1}}$ range $0.5$--$0.9$) makes
fixed-margin training suboptimal.
Lion provides the largest single-step lift on SCIDOCS ($+4.1\%$) and
FiQA ($+1.0\%$) when added last, consistent with its advantage on
clean pseudo-label distributions where sign-based updates \citep{chen2023symbolic} converge
faster than SGD under a small step budget.
The full DART model achieves the highest average gain,
confirming that the three components are complementary rather than
redundant.

%% ============================================================
\section{Analysis}
\label{sec:analysis}
%% ============================================================

\subsection{What Does \(W\) Learn?}

We denote the update of the transformation matrix after processing a query as
\begin{equation}
\Delta W = W^* - I,
\end{equation}
where $W^*$ is the matrix obtained after online adaptation. The Frobenius norm $\|\Delta W\|_F$ measures the magnitude of the update.

We analyze the average update over 50 randomly sampled test queries:
\begin{equation}
\overline{\Delta W} = \frac{1}{n}\sum_{i=1}^{n} \Delta W_i.
\end{equation}
Its singular value decomposition $\overline{\Delta W} = U \Sigma V^\top$ reveals a clear low-rank structure. The singular value matrix $\Sigma$ (only the largest few entries shown) is
\begin{equation}
\Sigma = \begin{bmatrix}
0.0116 & 0 & 0 & \cdots \\
0 & 0.0083 & 0 & \cdots \\
0 & 0 & 0.0051 & \cdots \\
\vdots & \vdots & \vdots & \ddots
\end{bmatrix},
\end{equation}
where $\sigma_1 = 0.0116$ accounts for 19.4\% of the total variation, and the top three singular values cumulatively explain 28.4\% of the variance. In contrast, a random matrix with the same Frobenius norm would exhibit a much flatter spectrum, with each of the 384 singular values approximately $0.0010$. This low-rank structure indicates that DART learns \emph{structured} semantic adjustments---rotating the scoring geometry in a small number of task-relevant directions---rather than making arbitrary perturbations.

\subsection{How Much Does \(W\) Change?}

The Frobenius norm of the identity matrix is
\begin{equation}
\|I\|_F = \sqrt{d} = \sqrt{384} \approx 19.6.
\end{equation}
The updates $\Delta W$ remain very small in comparison. Table~\ref{tab:update_stats} summarizes the distribution of $\|\Delta W\|_F$ across queries. The median update norm is only $0.095$, about 0.5\% of $\|I\|_F$, confirming that the regularization term successfully constrains adaptation. The distribution is right-skewed, reflecting the heterogeneity of query difficulty.

\subsection{How Does Query Affect \(W\) Change?}
\label{analysis-query}

Grouping queries by their top-1 retrieval similarity $s_{\text{top1}}$ reveals a basically monotonic relationship. Table~\ref{tab:difficulty} reports the average $\|\Delta W\|_F$ for four similarity intervals. Difficult queries (lower $s_{\text{top1}}$) receive larger updates, validating the adaptive margin heuristic: the model naturally allocates more adaptation capacity to queries that need it most.

\begin{table}[!t]
\centering
\small
\setlength{\tabcolsep}{2pt}
\caption{Average $\|\Delta W\|_F$ by $s_{\text{top1}}$ on NFCorpus.}
\label{tab:difficulty}
\begin{tabular}{lcccc}
\toprule
\textbf{Interval} & $[0.5, 0.6)$ & $[0.6, 0.7)$ & $[0.7, 0.8)$ & $[0.8, 1.0)$ \\
\midrule
Mean $\|\Delta W\|_F$ & 0.107 & 0.108 & 0.081 & 0.060 \\
\bottomrule
\end{tabular}
\end{table}

%% ============================================================
\section{Conclusion}
\label{sec:conclusion}
%% ============================================================

We presented DART, a zero-resource reranking framework that adapts a
bilinear scoring matrix at inference time using confidence-weighted
pseudo-labels derived directly from the dense retrieval ranking.
On six BEIR benchmarks, DART achieves a mean per-dataset relative
NDCG@10 gain of +2.1\% over the dense retrieval baseline
(under 10ms latency per query), demonstrating a powerful capability for zero-shot performance enhancement and cross-domain generalization.
The cross-query momentum mechanism (MetaInit and EMA) improves both
robustness and convergence speed across the query stream.
Interpretability analysis confirms that $W$ updates are conservative,
low-rank, and correlated with query difficulty.
Promising future directions include session-based retrieval and cross-lingual retrieval, where the semantic-lexical
gap is even more pronounced.

%% ============================================================
\section*{Limitations}
%% ============================================================

The dataset-adaptive optimizer selection strategy requires processing
50--100 queries with both SGD and Lion before committing to one.
In practice, SGD is the safer default: its conservative updates yield
neutral-to-positive gains across all tested datasets.
Lion is more aggressive---it can deliver larger improvements on clean
pseudo-label distributions (e.g., NFCorpus, FiQA) but may produce
zero or negative gains on noisier settings such as TREC-COVID.
In truly single-pass or streaming deployments where this warm-up is
undesirable, we recommend defaulting to SGD.

A second limitation concerns scalability to larger encoders.
DART currently optimizes a full $d \times d$ matrix $W$, whose memory
and per-query computation grow quadratically with the embedding
dimension.
For encoders with $d \ge 768$ (e.g., large BERT-family models), a
low-rank parameterization $W = I + AB^\top$ with $A, B \in
\mathbb{R}^{d \times r}$ and $r \ll d$ would substantially reduce
overhead without sacrificing performance; we
leave this extension to future work.

\section*{Acknowledgments}
We would like to thank the anonymous reviewers for their insightful comments and constructive suggestions, which helped improve the quality of this paper.

\bibliography{custom}

%% ============================================================
\appendix

\end{document}

%% file: figure1.tex
\begin{tikzpicture}

%% ===================== 坐标系网格 (Grid = 4.8 x 1.6) ===================== %%

%% Col 1: Input (x=0)
\node[hbox] (query)    at (0,  0)    {Query $q$};
\node[box]  (encoder)  at (0, -1.6)  {Query Encoder $\phi$};
\node[box]  (retrieve) at (0, -3.2)  {Top-$K$ Retrieval\\[-2pt]{\footnotesize cosine similarity}};

%% Col 2: Pseudo Labels (x=4.8)
\node[box]  (pos) at (4.8, 0) {Pseudo-Positives\\[-2pt]{\footnotesize top $n_+$ docs}};
\node[box]  (neg) at (4.8, -1.6) {Pseudo-Negatives\\[-2pt]{\footnotesize bottom $n_-$ docs}};
\node[box]  (cw)  at (4.8, -3.2) {Confidence Weights\\[-2pt]$w^{+},\,w^{-}$};

%% Col 3: Online Optimisation (x=9.6)
\node[box]  (winit) at (9.6, 0) {Init $W \leftarrow W_{\mathrm{meta}}$\\[-2pt]{\footnotesize (identity at start)}};
\node[lbox] (loss)  at (9.6, -1.6) {Margin Loss $\mathcal{L}$\\[-2pt]{\scriptsize $+\,\ell_2$\;reg.\;$\|W{-}I\|_F^2$}};
\node[box]  (opt)   at (9.6, -3.2) {Optimizer Step\\[-2pt]{\footnotesize SGD / Lion}};

%% Col 4: Momentum & Output (x=14.4)
\node[box]  (meta)   at (14.4, 0) {MetaInit Update\\[-2pt]$W_{\mathrm{meta}}$};
\node[box]  (ema)    at (14.4, -1.6) {EMA Update\\[-2pt]$W_{\mathrm{ema}}$};
\node[obox] (output) at (14.4, -3.2) {Reranked List\\[-2pt]{\footnotesize $\phi(q)^\top W_{\mathrm{ema}}\,\psi(d)$}};

%% ===================== 优雅连线 (Routing) ===================== %%

%% Col 1 内部连线
\draw[arr] (query)   -- (encoder);
\draw[arr] (encoder) -- (retrieve);

%% Col 1 -> Col 2 (Retrieve 分发)
\coordinate (ret_out) at ($(retrieve.east) + (0.5,0)$);

% Top 分支：向上折线
\draw[arr, rounded corners=4pt] (retrieve.east) -- (ret_out) |- (pos.west);
\node[above, font=\scriptsize\sffamily\color{Slate}] at ($(ret_out |- pos.west)!0.5!(pos.west)$) {top};

% Bottom 分支：改为折线，保持对称
\draw[arr, rounded corners=4pt] (retrieve.east) -- (ret_out) |- (neg.west);
\node[below, font=\scriptsize\sffamily\color{Slate}] at ($(ret_out |- neg.west)!0.5!(neg.west)$) {bottom};

%% Col 2 内部连线 (无交叉路由：pos 从右侧平滑绕过 neg)
\draw[arr, rounded corners=4pt] (pos.east) -- ++(0.4,0) |- (cw.10);
\draw[arr] (neg.south) -- (cw.north);

%% Col 2 -> Col 3
% cw.east 向右延伸 0.6 单位后，折线连接到 loss 的左侧中心 (west)
\draw[arr, rounded corners=4pt] (cw.east) -- ++(0.6,0) |- (loss.west);

\draw[arr] (winit.south) -- (loss.north);
\draw[arr] (loss.south)  -- (opt.north);

%% 对应的 Loop 连线微调 (确保不重叠)
% 将循环反馈线接入 loss 的右侧中心 (east)，形成左右对称的视觉感
\draw[arr, rounded corners=4pt] (opt.east) -- ++(0.5,0) |- (loss.east);
\node[font=\scriptsize\bfseries\sffamily, text=Slate, right] at ($(opt.east) + (-0.8, 0.8)$) {\texttt{steps}$\times$};

% %% Col 3 -> Col 4 (按高度分配外延通道，完美避免交叉)
% \draw[arr, rounded corners=4pt] (opt.east) -- ++(0.7,0) |- (meta.west);
% \draw[arr, rounded corners=4pt] (opt.east) -- ++(1.0,0) |- (ema.west);

%% Col 4 内部连线
\draw[arr] (ema.south) -- (output.north);

%% MetaInit 顶层反馈线
\draw[darr, rounded corners=4pt] (meta.west) -- (winit.east);
\node[font=\scriptsize\sffamily, text=Slate, inner xsep=4pt, inner ysep=2pt] 
      at ($(meta.north)!0.5!(winit.north) + (0.05, -0.2)$) {next query};

%% ===================== 背景与标签 ===================== %%

\begin{scope}[on background layer]
  % 统一设定面板的 padding 和质感投影
  \tikzset{
    panel/.style={
      fill=Bg, rounded corners=8pt, line width=1pt,
      inner xsep=12pt, inner ysep=15pt,
      drop shadow={opacity=0.05, shadow xshift=2.5pt, shadow yshift=-2.5pt}
    }
  }
  
  \node[panel, draw=Blue!25,  fit=(query)(encoder)(retrieve)] (pA) {};
  \node[panel, draw=Blue!25,  fit=(pos)(neg)(cw)]             (pB) {};
  \node[panel, draw=Red!25,   fit=(winit)(loss)(opt)]         (pC) {};
  \node[panel, draw=Green!30, fit=(meta)(ema)(output)]        (pD) {};
\end{scope}

%% 面板标签 (美化为右上角的小 Label)
\node[tag=Navy, anchor=south west]           at ($(pA.north west) + (4pt, 0)$) {Input};
\node[tag=Navy, anchor=south west]           at ($(pB.north west) + (4pt, 0)$) {Pseudo Labels};
\node[tag=Red!75!black, anchor=south west]   at ($(pC.north west) + (4pt, 0)$) {Online Optimization};
\node[tag=Green!55!black, anchor=south west] at ($(pD.north west) + (4pt, 0)$) {Momentum \& Output};

\end{tikzpicture}

%% file: custom.bib
@book{lin2022pretrained,
  title={Pretrained transformers for text ranking: Bert and beyond},
  author={Lin, Jimmy and Nogueira, Rodrigo and Yates, Andrew},
  year={2022},
  publisher={Springer Nature}
}

@article{guo2020deep,
  title={A deep look into neural ranking models for information retrieval},
  author={Guo, Jiafeng and Fan, Yixing and Pang, Liang and Yang, Liu and Ai, Qingyao and Zamani, Hamed and Wu, Chen and Croft, W Bruce and Cheng, Xueqi},
  journal={Information Processing \& Management},
  volume={57},
  number={6},
  pages={102067},
  year={2020},
  publisher={Elsevier}
}

@inproceedings{karpukhin2020dense,
  title={Dense passage retrieval for open-domain question answering},
  author={Karpukhin, Vladimir and Oguz, Barlas and Min, Sewon and Lewis, Patrick and Wu, Ledell and Edunov, Sergey and Chen, Danqi and Yih, Wen-tau},
  booktitle={Proceedings of the 2020 conference on empirical methods in natural language processing (EMNLP)},
  pages={6769--6781},
  year={2020}
}

@inproceedings{reimers2019sentence,
  title={Sentence-bert: Sentence embeddings using siamese bert-networks},
  author={Reimers, Nils and Gurevych, Iryna},
  booktitle={Proceedings of the 2019 conference on empirical methods in natural language processing and the 9th international joint conference on natural language processing (EMNLP-IJCNLP)},
  pages={3982--3992},
  year={2019}
}

@inproceedings{xiao2024c,
  title={C-pack: Packed resources for general chinese embeddings},
  author={Xiao, Shitao and Liu, Zheng and Zhang, Peitian and Muennighoff, Niklas and Lian, Defu and Nie, Jian-Yun},
  booktitle={Proceedings of the 47th international ACM SIGIR conference on research and development in information retrieval},
  pages={641--649},
  year={2024}
}

@article{nogueira2019passage,
  title={Passage Re-ranking with BERT},
  author={Nogueira, Rodrigo and Cho, Kyunghyun},
  journal={arXiv preprint arXiv:1901.04085},
  year={2019}
}

@inproceedings{nogueira2020document,
  title={Document ranking with a pretrained sequence-to-sequence model},
  author={Nogueira, Rodrigo and Jiang, Zhiying and Pradeep, Ronak and Lin, Jimmy},
  booktitle={Findings of the association for computational linguistics: EMNLP 2020},
  pages={708--718},
  year={2020}
}

@inproceedings{santhanam2022colbertv2,
  title={Colbertv2: Effective and efficient retrieval via lightweight late interaction},
  author={Santhanam, Keshav and Khattab, Omar and Saad-Falcon, Jon and Potts, Christopher and Zaharia, Matei},
  booktitle={Proceedings of the 2022 Conference of the North American Chapter of the Association for Computational Linguistics: Human Language Technologies},
  pages={3715--3734},
  year={2022}
}

@inproceedings{sun2023chatgpt,
  title={Is ChatGPT good at search? investigating large language models as re-ranking agents},
  author={Sun, Weiwei and Yan, Lingyong and Ma, Xinyu and Wang, Shuaiqiang and Ren, Pengjie and Chen, Zhumin and Yin, Dawei and Ren, Zhaochun},
  booktitle={Proceedings of the 2023 conference on empirical methods in natural language processing},
  pages={14918--14937},
  year={2023}
}

@article{pradeep2023rankvicuna,
  title={Rankvicuna: Zero-shot listwise document reranking with open-source large language models},
  author={Pradeep, Ronak and Sharifymoghaddam, Sahel and Lin, Jimmy},
  journal={arXiv preprint arXiv:2309.15088},
  year={2023}
}

@article{weller2025rank1,
  title={Rank1: Test-time compute for reranking in information retrieval},
  author={Weller, Orion and Ricci, Kathryn and Yang, Eugene and Yates, Andrew and Lawrie, Dawn and Van Durme, Benjamin},
  journal={arXiv preprint arXiv:2502.18418},
  year={2025}
}

@inproceedings{devlin2019bert,
  title={Bert: Pre-training of deep bidirectional transformers for language understanding},
  author={Devlin, Jacob and Chang, Ming-Wei and Lee, Kenton and Toutanova, Kristina},
  booktitle={Proceedings of the 2019 conference of the North American chapter of the association for computational linguistics: human language technologies, volume 1 (long and short papers)},
  pages={4171--4186},
  year={2019}
}

@inproceedings{wang2022gpl,
  title={GPL: Generative pseudo labeling for unsupervised domain adaptation of dense retrieval},
  author={Wang, Kexin and Thakur, Nandan and Reimers, Nils and Gurevych, Iryna},
  booktitle={Proceedings of the 2022 conference of the North American chapter of the association for computational linguistics: human language technologies},
  pages={2345--2360},
  year={2022}
}

@inproceedings{zhuang2023augmenting,
  title={Augmenting passage representations with query generation for enhanced cross-lingual dense retrieval},
  author={Zhuang, Shengyao and Shou, Linjun and Zuccon, Guido},
  booktitle={Proceedings of the 46th International ACM SIGIR Conference on Research and Development in Information Retrieval},
  pages={1827--1832},
  year={2023}
}

@article{meng2022unsupervised,
  title={Unsupervised dense retrieval deserves better positive pairs: Scalable augmentation with query extraction and generation},
  author={Meng, Rui and Liu, Ye and Yavuz, Semih and Agarwal, Divyansh and Tu, Lifu and Yu, Ning and Zhang, Jianguo and Bhat, Meghana and Zhou, Yingbo},
  journal={arXiv preprint arXiv:2212.08841},
  year={2022}
}

@inproceedings{saad2023udapdr,
  title={UDAPDR: unsupervised domain adaptation via LLM prompting and distillation of rerankers},
  author={Saad-Falcon, Jon and Khattab, Omar and Santhanam, Keshav and Florian, Radu and Franz, Martin and Roukos, Salim and Sil, Avirup and Sultan, Md and Potts, Christopher},
  booktitle={Proceedings of the 2023 conference on empirical methods in natural language processing},
  pages={11265--11279},
  year={2023}
}

@inproceedings{lavrenko2017relevance,
  title={Relevance-based language models},
  author={Lavrenko, Victor and Croft, W Bruce},
  booktitle={ACM SIGIR Forum},
  volume={51},
  number={2},
  pages={260--267},
  year={2017},
  organization={ACM New York, NY, USA}
}

@article{li2023pseudo,
  title={Pseudo relevance feedback with deep language models and dense retrievers: Successes and pitfalls},
  author={Li, Hang and Mourad, Ahmed and Zhuang, Shengyao and Koopman, Bevan and Zuccon, Guido},
  journal={ACM Transactions on Information Systems},
  volume={41},
  number={3},
  pages={1--40},
  year={2023},
  publisher={ACM New York, NY}
}

@article{wang2023colbert,
  title={ColBERT-PRF: Semantic pseudo-relevance feedback for dense passage and document retrieval},
  author={Wang, Xiao and Macdonald, Craig and Tonellotto, Nicola and Ounis, Iadh},
  journal={ACM Transactions on the Web},
  volume={17},
  number={1},
  pages={1--39},
  year={2023},
  publisher={ACM New York, NY}
}

@article{li2025pseudo,
  title={Pseudo Relevance Feedback is Enough to Close the Gap Between Small and Large Dense Retrieval Models},
  author={Li, Hang and Wang, Xiao and Koopman, Bevan and Zuccon, Guido},
  journal={arXiv preprint arXiv:2503.14887},
  year={2025}
}

@article{wang2022text,
  title={Text embeddings by weakly-supervised contrastive pre-training},
  author={Wang, Liang and Yang, Nan and Huang, Xiaolong and Jiao, Binxing and Yang, Linjun and Jiang, Daxin and Majumder, Rangan and Wei, Furu},
  journal={arXiv preprint arXiv:2212.03533},
  year={2022}
}

@inproceedings{sun2020test,
  title={Test-time training with self-supervision for generalization under distribution shifts},
  author={Sun, Yu and Wang, Xiaolong and Liu, Zhuang and Miller, John and Efros, Alexei and Hardt, Moritz},
  booktitle={International conference on machine learning},
  pages={9229--9248},
  year={2020},
  organization={PMLR}
}

@article{liu2021ttt++,
  title={Ttt++: When does self-supervised test-time training fail or thrive?},
  author={Liu, Yuejiang and Kothari, Parth and Van Delft, Bastien and Bellot-Gurlet, Baptiste and Mordan, Taylor and Alahi, Alexandre},
  journal={Advances in Neural Information Processing Systems},
  volume={34},
  pages={21808--21820},
  year={2021}
}

@article{akyurek2024surprising,
  title={The surprising effectiveness of test-time training for few-shot learning},
  author={Aky{\"u}rek, Ekin and Damani, Mehul and Zweiger, Adam and Qiu, Linlu and Guo, Han and Pari, Jyothish and Kim, Yoon and Andreas, Jacob},
  journal={arXiv preprint arXiv:2411.07279},
  year={2024}
}

@article{yu2023metamath,
  title={Metamath: Bootstrap your own mathematical questions for large language models},
  author={Yu, Longhui and Jiang, Weisen and Shi, Han and Yu, Jincheng and Liu, Zhengying and Zhang, Yu and Kwok, James T and Li, Zhenguo and Weller, Adrian and Liu, Weiyang},
  journal={arXiv preprint arXiv:2309.12284},
  year={2023}
}

@article{chen2023symbolic,
  title={Symbolic discovery of optimization algorithms},
  author={Chen, Xiangning and Liang, Chen and Huang, Da and Real, Esteban and Wang, Kaiyuan and Pham, Hieu and Dong, Xuanyi and Luong, Thang and Hsieh, Cho-Jui and Lu, Yifeng and others},
  journal={Advances in neural information processing systems},
  volume={36},
  pages={49205--49233},
  year={2023}
}

@article{thakur2021beir,
  title={Beir: A heterogenous benchmark for zero-shot evaluation of information retrieval models},
  author={Thakur, Nandan and Reimers, Nils and R{\"u}ckl{\'e}, Andreas and Srivastava, Abhishek and Gurevych, Iryna},
  journal={arXiv preprint arXiv:2104.08663},
  year={2021}
}

@inproceedings{abdallah2025asrank,
  title={Asrank: Zero-shot re-ranking with answer scent for document retrieval},
  author={Abdallah, Abdelrahman and Mozafari, Jamshid and Piryani, Bhawna and Jatowt, Adam},
  booktitle={Findings of the Association for Computational Linguistics: NAACL 2025},
  pages={2950--2970},
  year={2025}
}

@article{chen2024attention,
  title={Attention in large language models yields efficient zero-shot re-rankers},
  author={Chen, Shijie and Guti{\'e}rrez, Bernal Jim{\'e}nez and Su, Yu},
  journal={arXiv preprint arXiv:2410.02642},
  year={2024}
}

@article{huang2024instupr,
  title={InstUPR: Instruction-based unsupervised passage reranking with large language models},
  author={Huang, Chao-Wei and Chen, Yun-Nung},
  journal={arXiv preprint arXiv:2403.16435},
  year={2024}
}

@article{johnson2019billion,
  title={Billion-scale similarity search with GPUs},
  author={Johnson, Jeff and Douze, Matthijs and J{\'e}gou, Herv{\'e}},
  journal={IEEE transactions on big data},
  volume={7},
  number={3},
  pages={535--547},
  year={2019},
  publisher={IEEE}
}

@book{robertson2009probabilistic,
  title={The probabilistic relevance framework: BM25 and beyond},
  author={Robertson, Stephen and Zaragoza, Hugo},
  volume={4},
  year={2009},
  publisher={Now Publishers Inc}
}

@inproceedings{khattab2020colbert,
  title={Colbert: Efficient and effective passage search via contextualized late interaction over bert},
  author={Khattab, Omar and Zaharia, Matei},
  booktitle={Proceedings of the 43rd International ACM SIGIR conference on research and development in Information Retrieval},
  pages={39--48},
  year={2020}
}

@inproceedings{xin2022zero,
  title={Zero-shot dense retrieval with momentum adversarial domain invariant representations},
  author={Xin, Ji and Xiong, Chenyan and Srinivasan, Ashwin and Sharma, Ankita and Jose, Damien and Bennett, Paul},
  booktitle={Findings of the Association for Computational Linguistics: ACL 2022},
  pages={4008--4020},
  year={2022}
}

@article{xiong2020approximate,
  title={Approximate nearest neighbor negative contrastive learning for dense text retrieval},
  author={Xiong, Lee and Xiong, Chenyan and Li, Ye and Tang, Kwok-Fung and Liu, Jialin and Bennett, Paul and Ahmed, Junaid and Overwijk, Arnold},
  journal={arXiv preprint arXiv:2007.00808},
  year={2020}
}

@inproceedings{hofst2021efficiently,
author = {Hofst\"{a}tter, Sebastian and Lin, Sheng-Chieh and Yang, Jheng-Hong and Lin, Jimmy and Hanbury, Allan},
title = {Efficiently Teaching an Effective Dense Retriever with Balanced Topic Aware Sampling},
year = {2021},
isbn = {9781450380379},
publisher = {Association for Computing Machinery},
address = {New York, NY, USA},
url = {https://doi.org/10.1145/3404835.3462891},
doi = {10.1145/3404835.3462891},
booktitle = {Proceedings of the 44th International ACM SIGIR Conference on Research and Development in Information Retrieval},
pages = {113–122},
numpages = {10},
location = {Virtual Event, Canada},
series = {SIGIR '21}
}

@article{Izacard2021UnsupervisedDI,
  title={Unsupervised Dense Information Retrieval with Contrastive Learning},
  author={Gautier Izacard and Mathilde Caron and Lucas Hosseini and Sebastian Riedel and Piotr Bojanowski and Armand Joulin and Edouard Grave},
  journal={Trans. Mach. Learn. Res.},
  year={2021},
  volume={2022},
  url={https://api.semanticscholar.org/CorpusID:249097975}
}

@article{DeepSeekAI2025DeepSeekR1IR,
  title={DeepSeek-R1 incentivizes reasoning in LLMs through reinforcement learning},
  author={DeepSeek-AI and Daya Guo and Dejian Yang and Haowei Zhang and Jun-Mei Song and Ruoyu Zhang and Runxin Xu and Qihao Zhu and Shirong Ma and Peiyi Wang and Xiaoling Bi and Xiaokang Zhang and Xingkai Yu and Yu Wu and Z. F. Wu and Zhibin Gou and Zhihong Shao and Zhuoshu Li and Ziyi Gao and Aixin Liu and Bing Xue and Bing-Li Wang and Bochao Wu and Bei Feng and Chengda Lu and Chenggang Zhao and Chengqi Deng and Chenyu Zhang and Chong Ruan and Damai Dai and Deli Chen and Dong-Li Ji and Erhang Li and Fangyun Lin and Fucong Dai and Fuli Luo and Guangbo Hao and Guanting Chen and Guowei Li and H. Zhang and Han Bao and Hanwei Xu and Haocheng Wang and Honghui Ding and Huajian Xin and Huazuo Gao and Hui Qu and Hui Li and Jianzhong Guo and Jiashi Li and Jiawei Wang and JingChang Chen and Jingyang Yuan and Junjie Qiu and Junlong Li and Jiong Cai and Jiaqi Ni and Jian Liang and Jin Chen and Kai Dong and Kai Hu and Kaige Gao and Kang Guan and Kexin Huang and Kuai Yu and Lean Wang and Lecong Zhang and Liang Zhao and Litong Wang and Liyue Zhang and Lei Xu and Leyi Xia and Mingchuan Zhang and Minghua Zhang and M. Tang and Meng Li and Miaojun Wang and Mingming Li and Ning Tian and Panpan Huang and Peng Zhang and Qiancheng Wang and Qinyu Chen and Qiushi Du and Ruiqi Ge and Ruisong Zhang and Ruizhe Pan and Runji Wang and R. J. Chen and Ruiqi Jin and Ruyi Chen and Shanghao Lu and Shangyan Zhou and Shanhuang Chen and Shengfeng Ye and Shiyu Wang and Shuiping Yu and Shunfeng Zhou and Shuting Pan and S. S. Li and Shuang Zhou and Shao-Kang Wu and Tao Yun and Tian Pei and Tianyu Sun and T. Wang and Wangding Zeng and Wanjia Zhao and Wen Liu and Wenfeng Liang and Wenjun Gao and Wen-Xia Yu and Wentao Zhang and Wangding Xiao and Wei An and Xiaodong Liu and Xiaohan Wang and Xiaokang Chen and Xiaotao Nie and Xin Cheng and Xin Liu and Xin Xie and Xingchao Liu and Xinyu Yang and Xinyuan Li and Xuecheng Su and Xuheng Lin and X. Q. Li and Xiangyu Jin and Xi-Cheng Shen and Xiaosha Chen and Xiaowen Sun and Xiaoxiang Wang and Xinnan Song and Xinyi Zhou and Xianzu Wang and Xinxia Shan and Y. K. Li and Y. Q. Wang and Y. X. Wei and Yang Zhang and Yanhong Xu and Yao Li and Yao Zhao and Yaofeng Sun and Yaohui Wang and Yi Yu and Yichao Zhang and Yifan Shi and Yi Xiong and Ying He and Yishi Piao and Yisong Wang and Yixuan Tan and Yiyang Ma and Yiyuan Liu and Yongqiang Guo and Yuan Ou and Yuduan Wang and Yue Gong and Yu-Jing Zou and Yujia He and Yunfan Xiong and Yu-Wei Luo and Yu-mei You and Yuxuan Liu and Yuyang Zhou and Y. X. Zhu and Yanping Huang and Yao Li and Yi Zheng and Yuchen Zhu and Yunxiang Ma and Ying Tang and Yukun Zha and Yuting Yan and Zehui Ren and Zehui Ren and Zhangli Sha and Zhe Fu and Zhean Xu and Zhenda Xie and Zhen-guo Zhang and Zhewen Hao and Zhicheng Ma and Zhigang Yan and Zhiyu Wu and Zihui Gu and Zijia Zhu and Zijun Liu and Zi-An Li and Ziwei Xie and Ziyang Song and Zizheng Pan and Zhen Huang and Zhipeng Xu and Zhongyu Zhang and Zhen Zhang},
  journal={Nature},
  year={2025},
  volume={645},
  pages={633 - 638},
  url={https://api.semanticscholar.org/CorpusID:275789950}
}
